Dolgopolov Stanislav
dolgopolov-s@list.ru

**Pairing of valence electrons as necessary condition for energy minimization in a crystal**

Pairing of valence electrons can lead to energy minimization of a crystal. It can be proved by use of representation of the valence electrons as plane waves in periodic potential of the crystal.

1. **Valence electrons in crystals as standing wave packets.**

According to the Bloch theorem an electron wave function in a periodic potential of a crystal is a plane wave running along axis *x* :

$$\Psi'(x,t) = A \cdot u(x) \cdot e^{(-i\omega t - ikx)} \qquad (1.1)$$

Where :
*ω*    cyclic frequency
*k*    wave number
*A*    amplitude of the wave packet. Generally *A* can be a function of *x* and *t*
*u(x)*    a function of *x* with a period equal to the lattice constant ([1]). The function *u(x)* represents approximately a solution of the Schrödinger equation for electron in a 1D box ([2]).

On the other hand an electron wave has a reflection on ions of lattice, what can lead to a full reflection of the electron. Then the electron becomes a standing wave with a zero momentum.
The condition of the full reflection is the Bregg-condition ([3]):

$$n\lambda = 2 \cdot R_0 \qquad (1.2)$$

Where :
*n*    integer
*λ*    length of the electron wave Ψ(x, t)
$R_0$    lattice constant.

We note that *λ* in eq. (1.2) is order of magnitude of lattice constant, what corresponds to the electron energy order of magnitude of Fermi energy.
A standing wave can be described as a sum of a direct wave and a reflected wave running in opposite direction :

$$\Psi(x,t) = \frac{1}{\sqrt{2}} A \cdot u(x) \cdot e^{(-i\omega t - ikx)} + \frac{1}{\sqrt{2}} A \cdot u(x) \cdot e^{(-i\omega t + ikx)} = A \cdot u(x) \cdot e^{-i\omega t} \sqrt{2} \cdot \cos kx \qquad (1.3)$$

If the reflection coefficient on every point of lattice is not zero then the amplitude *A* of the running wave decreases on each ion and becomes negligible after a number of points of lattice. The number can be very great in case of a good conductor; in spite of that one can say that the location area of a standing wave has a finite size.
Momentum of all valence electrons in a crystal are in a range, consequently the wave lengths are also in a range. We investigate a wave length $\lambda = 2 \cdot R_0$ corresponding to the Bregg-condition (1.2) at *n=1*.

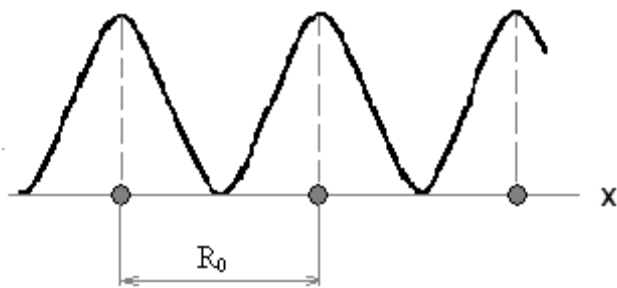

Figure 1. Electron wave density $|\Psi(x)|^2$ at $\lambda = 2 \cdot R_0$

The wave density $|\Psi(x,t)|^2$ of the cosine in eq. (1.3) at $\lambda = 2 \cdot R_0$ is shown in figure 1. Peaks of $|\Psi(x,t)|^2$ coincide with the ions of lattice, what corresponds to the minimal average distance of an electron to ions of lattice. This leads to a maximal attraction of the electron to ions.

For energy calculation we assume that electron 1 *($e_1$)* is distributed in an area of crystal, where the amplitude of the standing wave is not negligible. All negative and positive charges of the crystal outside the area $e_1$ are compensated, therefore we take into account only ions and electrons located inside the area of electron 1.
The potential energy of electron 1 contains 2 parts :

1. Energy of attraction $e_1$ to the community of positive ions located in the $e_1$ area. We designate it as $P(e_1,I)$.

$$\langle P(e_1,I)\rangle = -\lim_{\Delta V_i \to 0} \sum_{i=1}^{n_1} \sum_{k=1}^{n} |\Psi_1(\vec{r_i})|^2 \cdot \Delta V_i \cdot \frac{e \cdot q_k}{|\vec{r_i} - \vec{r_k}|} \qquad (1.4)$$

Where :

$\Delta V_i$    volume element of the area of electron 1

$\vec{r_i}$    radius-vector of element $\Delta V_i$

$\Psi_1(\vec{r_i})$    wave function of electron 1 in volume element $\Delta V_i$

$q_k$    charge of one ion located in area of electron 1 ( $q_k$ = +e for one-valent atoms)

$\vec{r_k}$    radius-vector of the charge $q_k$

$n$    number of ions located in area of electron 1

$n_1$    number of volume elements $\Delta V_i$ in area of electron 1

$e$    elementary charge

A minimal value $\langle P(e_1,I)\rangle$ calculated by eq. (1.4) and a maximal attraction occurs when the electron wave is distributed as in figure 1. This $|\Psi(x,t)|^2$ has a minimal average distance between electron 1 and ions of lattice. If $\lambda \neq 2 \cdot R_0$ or the wave is running, then the electron density gets more into the space between ions, the average distance of $e_1$ to ions increases, the attraction weakens.

2. Energy of repulsion of electron 1 from the community of other electrons located in area of electron 1. We designate it as $P(e_1,e)$.

$$\langle P(e_1,e)\rangle = \lim_{\substack{\Delta V_i \to 0 \\ \Delta V_k \to 0}} \sum_{\alpha=2}^{m} \sum_{i=1}^{n_1} \sum_{k=1}^{n_\alpha} e^2 \frac{|\Psi_1(\vec{r_i})|^2 \cdot \Delta V_i \cdot |\Psi_\alpha(\vec{r_k})|^2 \cdot \Delta V_k}{|\vec{r_i} - \vec{r_k}|} \qquad (1.5)$$

Where :

$\Delta V_k$    volume element of area of electron α (α is the order number of electrons)

$\vec{r_k}$    radius-vector of element $\Delta V_k$ of electron α

$\Psi_\alpha(\vec{r_k})$    wave function of electron α in element $\Delta V_k$

$n_\alpha$    number of volume elements $\Delta V_k$ in area of electron α

$m$    number of electrons in area of electron 1 (we assume the number of electrons is equal to the number of ions).

A maximal repulsion $\langle P(e_1,e)\rangle$ occurs when all electrons are distributed as in figure 1 at $\lambda = 2 \cdot R_0$. Calculations made by use of eq. (1.4) and (1.5) show: in spite of the maximal electron repulsion the case in fig. 1 is preferable in comparison with cases when $\lambda \neq 2 \cdot R_0$ or when the waves are running. The matter is that by all configurations (except as in figure1) the electron density loses the minimal average distance to ions, therefore $P(e_1,I)$ grows and $P(e_1,e)$ decreases, but $P(e_1,I)$ grows faster than $P(e_1,e)$ decreases, consequently the total potential energy $\langle P(e_1,I)\rangle + \langle P(e_1,e)\rangle$ increases. For density configurations ( $\lambda \neq 2 \cdot R_0$ or waves are running) the gains from the reducing of repulsion are less than the loss from the reducing of attraction.

In case of figure 1 a spatial divergence of the electron densities around each ion leads to the increasing potential energy of electrons. This fact has a simple explanation: in figure 1 the average distance of the 3-dimensionally distributed electron 1 to the point-like ions is **shorter** than the average distance of the 3-dimensional electron 1 to all other coinciding 3-dimensional electrons. As a result by every displacement of the electron densities from the position in figure 1 the decreasing of attraction is greater than the decreasing of repulsion. Thus a convergence of the wave densities in fig. 1 is more probable than a divergence. This fact is clearly calculable by use of eq. (1.4) and (1.5).

We note that density configurations of $u(x)$ are possible, where the average distance of each electron to ions is **longer** than the average distance of each electron to other electrons. In this case a spatial divergence of the electron densities can be preferable and a spatial convergence is less probable.

In the case of figure1 (where the peaks of the wave densities coincide with the ions of lattice) all calculations show :

$$\langle P_{e1} \rangle = \langle P(e_1, I) \rangle + \langle P(e_1, e) \rangle = MINIMUM < 0 \quad at \ \lambda = 2 \cdot R_0 \qquad (1.6)$$

Thus one can designate the state in fig. 1 as a potential well. It is easy to show that the kinetic energies of a running wave and of a related standing wave are equal; consequently the full energy of electron in fig. 1 is also minimal. This means that the system ions-electron loses some energy during the transition into the potential well. Therefore the state at $\lambda = 2 \cdot R_0$ can exist.

We note that according to the calculations made by use of eq. (1.4) and (1.5) the well depth goes to zero if the $e_1$ area is infinite. As a result the potential well depends on the reflection coefficient of valence electrons on the ions of lattice; a greater reflection coefficient makes the $e_1$ area more space saving, consequently the well depth increases.

2. **Minimization of energy in a crystal.**

The total energy of a valence electron in a crystal contains two parts ([4]) :

1. Potential and kinetic energies of interaction with lattice. The part depends on the term $u(x)$ in eq. (1.1). We designate the part as $E_u = P_u + K_u$. The energy $E_u$ almost does not depend on the wave length $\lambda$ ([5]).

2. Kinetic energy of the electron motion in the crystal. The part depends on the term $e^{(-i\omega t - ikx)}$ in eq. (1.1). We designate the part as $E_\lambda$. The energy $E_\lambda$ can be considered as kinetic energy of a free electron in a constant potential ([6]):

$$E_\lambda = \frac{h^2}{\lambda^2 \cdot 2 \cdot m} \qquad (2.1)$$

Where $m$ is the mass of electron.

Due to the exclusion principle the energy $E_\lambda$ of all valence electrons is in a range from zero to the Fermi energy $E_F$ ([7]). The value $\lambda = 2 \cdot R_0$ corresponds to the Bregg-condition (1.2) for standing waves at $n=1$. At $\lambda = 2 \cdot R_0$ the energy $E_\lambda$ is :

$$E_{\lambda = 2R_0} = \frac{h^2}{R_0^2 \cdot 8 \cdot m} \qquad (2.2)$$

Since the location area of a standing wave has a finite size in space, the potential well at $\lambda = 2 \cdot R_0$ has also a finite size in space. Two identical electrons in one local potential well can build a pair like electrons in Helium atom (with the difference that in a crystal not one ion but a group of ions creates the potential well). Here the exchange interaction of two electrons has a critical importance. Due to a pairing with opposite spins an additional spatial convergence of electron densities occurs around every ion, and we already know : if the area of electrons around point-like ions converges, then the attraction of $e_1$ to ions grows faster than the repulsion of $e_1$ from other electrons grows. Consequently due to the opposite spins the potential well becomes deeper. (We note again that there are specific density configurations of $u(x)$ where a spatial divergence of electrons is preferable, then two spins can be parallel).

In the case $\lambda = 2 \cdot R_0$ in fig.1 two electrons and the group of associated ions build a quantum system with antiparallel spins of electrons. Two paired electrons can be considered as a boson, consequently the two electrons can be excluded from the energy statistic for single electrons, therefore two states with $\lambda = 2 \cdot R_0$ become vacant and can be occupied by two new valence electrons, which again can build a pair. Every new electron pair reduces the concentration of single electrons $N_e$ in the crystal. Since the Fermi energy is proportional to $N_e^{2/3}$ ([8]) the Fermi energy decreases. This process runs so long as the Fermi energy is greater than $E_{\lambda = 2R_0}$.

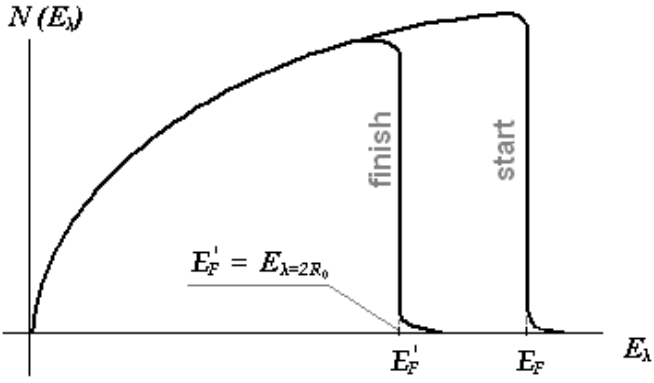

Figure 2.

Figure 2 shows the start value of the Fermi energy $E_F$ and the finish value $E_F^{'}$. When the finish value $E_F^{'} < E_{\lambda=2R_0}$ then there are no more single electrons with $\lambda = 2 \cdot R_0$, the pairing stops.

Using the equation $E_F(N_e)$ ([9]) and eq. (2.2) is easy to find the concentration of the electron pairs $N_p$:

$$N_p = \frac{1}{2}(N_e - N_e^{'}) = \frac{1}{2}\left(N_e - \frac{(E_{\lambda=2R_0} \cdot 8m)^{3/2} \cdot \pi}{3 \cdot h^3}\right) = \frac{1}{2}\left(N_e - \frac{\pi}{3 \cdot R_0^3}\right) \qquad (2.3)$$

Where:

$N_e$    initial concentration of single electrons when the Fermi energy is $E_F > E_{\lambda=2R_0}$

$N_e^{'}$    concentration of single electrons when the pairing process is finished and the Fermi energy is $E_F^{'} = E_{\lambda=2R_0}$.

One paired electron in the potential well gives a gain in the energy for the whole crystal, thus the pairing of all electrons having the energy from $E_F^{'} = E_{\lambda=2R_0}$ to $E_F > E_{\lambda=2R_0}$ reduces the total energy of the crystal. Without this pairing very few electrons could drop into the potential well because of the exclusion principle.

A gain in energy for one paired electron due to the transition into the potential well can be expressed as :

$$\Delta E = (P_s + K_s) - (P_u + K_u + E_\lambda)_{\lambda=2R_0} \qquad (2.4)$$

Where :

$P_s$    potential energy of electron in the potential well at $\lambda = 2 \cdot R_0$

$K_s$    kinetic energy of electron in the potential well at $\lambda = 2 \cdot R_0$.

Generally $(P_s + K_s) \neq (P_u + K_u)_{\lambda=2R_0}$ because $(P_u + K_u)_{\lambda=2R_0}$ is the energy of interaction of a single running electron with the lattice **before** transition into potential well; $(P_s + K_s)$ is the energy of an electron **after** pairing and after transition into potential well, where $E_\lambda$ has no sense.

If an electron in the potential well absorbs the energy (2.4), then it can leave the potential well and become again a running wave, then the pairing is lost. Therefore the thermal energy in the crystal can permanently destroy the electron pairs and the concentration of the pairs stays negligible.

References

1. Robert L. Sproull, Modern Physics, Moscow, Nauka, 1974

---

[1] [1, § 8.5, p.266]
[2] [1, § 8.5, p.266 and 1, § 5.4, p.151]
[3] [1, § 8.5, eq. (8.7)]
[4] [1, § 9.4, p.288]
[5] [1, § 9.4, p.289]
[6] [1, § 9.4, p.289]
[7] [1, § 9.4, p.290]
[8] [1, § 9.4, eq. (9.7)]
[9] [1, § 9.4, eq. (9.7)]